\documentclass[12pt]{article}
\input epsf.sty
\newcommand{\eps}{\epsilon}
\newcommand{\ra}{\rangle}
\newcommand{\la}{\langle}

\newcommand{\ve}{\varepsilon}
\newcommand{\tl}{\wt\lambda}

\newcommand{\BB}{{\cal B}}

\newcommand{\OO}{{\cal O}}

\newcommand{\wt}{\widetilde}

\newcommand{\TT}{{\cal T}}

\newcommand{\be}{\begin{equation}}
\newcommand{\ee}{\end{equation}}
\newcommand{\ben}{\begin{eqnarray}\displaystyle}
\newcommand{\een}{\end{eqnarray}}
\newcommand{\refb}[1]{(\ref{#1})}
\newcommand{\p}{\partial}
\newcommand{\sectiono}[1]{\section{#1}\setcounter{equation}{0}}

\begin{document}
{}~
\hfill\vbox{\hbox{hep-th/0203265}
}\break

\vskip .6cm

\centerline{\Large \bf
Tachyon Matter}

\medskip

\vspace*{4.0ex}

\centerline{\large \rm 
Ashoke Sen }

\vspace*{4.0ex}

\centerline{\large \it Harish-Chandra Research
Institute}

\centerline{\large \it  Chhatnag Road, Jhusi,
Allahabad 211019, INDIA}

\centerline {and}
\centerline{\large \it Department of Physics, Penn State University}

\centerline{\large \it University Park,
PA 16802, USA}

\centerline{E-mail: asen@thwgs.cern.ch, sen@mri.ernet.in}

\vspace*{5.0ex}

\centerline{\bf Abstract} \bigskip 

It is shown that classical decay of unstable D-branes in bosonic and
superstring theories produces pressureless gas with non-zero energy
density. The energy density is stored in the open string fields, even
though around the minimum of the tachyon potential there are no open
string degrees of freedom. We also give a description of this phenomenon
in an effective field theory.

\vfill \eject

\baselineskip=18pt

\tableofcontents

\sectiono{Introduction and Summary} \label{s1}

In a previous paper\cite{0203211} we developed a general formalism for
constructing classical time dependent solutions describing the rolling
tachyon field on unstable D-branes in bosonic and superstring theories.  
We also analysed the boundary state associated with these solutions in
bosonic string theory in some detail, and showed that the energy momentum
tensor approaches a finite limiting value if we push the system towards a
direction in which the tachyon potential has a minimum.

In this paper we carry out a detailed study of the energy momentum tensor
for rolling tachyon solution in D-branes in bosonic string theory, as well
as for unstable D-branes and brane-antibrane configurations in superstring 
theories. We conclude that in both
cases the asymptotic form of the energy momentum tensor for large time is
described by a pressureless gas with non-zero energy density.
The energy density is stored in the open string fields, even 
though around the minimum of the tachyon potential there are no open 
string degrees of freedom.

The paper is organised as follows. In section \ref{s2} we review the
procedure used for determining the energy momentum tensor in terms of
boundary states following \cite{9604091,9707068,9912275}, and 
determine 
the
expression for the energy momentum tensor for the rolling tachyon solution
in bosonic string theory. In sections \ref{s3} and \ref{s4} we generalize 
this procedure
to unstable D-branes in superstring theories. 
In both theories the energy density remains constant and the pressure 
approaches zero as the tachyon field rolls towards its minimum.
In section \ref{s5} we give 
a description of this phenomenon using an effective field theory.

It will be interesting to see if this kind of tachyon matter system can
play some role in cosmology; in particular if it could contribute to the
dark matter in the universe. Also interesting is the question as to
whether tachyon condensation in other kind of unstable brane systems ({\it
e.g.} branes at angles) could give rise to other kind of matter, with
different equation of state. Another relevant issue is the nature of
supersymmetry breaking induced by the tachyon matter. Since the total
energy of the tachyon matter is an adjustable parameter, determined by the
initial position and velocity of the tachyon, the associated supersymmetry
breaking scale will also be an adjustable parameter.

\sectiono{Analysis in Bosonic String Theory} \label{s2}

A boundary state $|\BB\ra$ associated with a D-brane system in bosonic 
string theory is a closed string state of ghost number 3, defined as 
follows.
For a closed string state $|\psi_c\ra$ of 
ghost number 3 (which could be obtained for example by applying the 
operator $(c_0-\bar c_0)$ on a physical closed string state of ghost 
number 2) $\la \BB|\psi_c\ra$ gives the one point function of the 
closed string vertex operator $\psi_c$ on the unit disk, with boundary 
condition appropriate to the particular D-brane system under 
consideration. The boundary state $|\BB\ra$ acts as a source for closed 
string fields; indeed it couples to the closed string field $|\Psi_c\ra$ 
of ghost number 2, satisfying $(b_0-\bar b_0)|\Psi_c\ra=0$\cite{9705241}, 
through a term proportional to $\la\BB|(c_0-\bar c_0)|\Psi_c\ra$. 
To 
linearized 
order the equation of motion of $|\Psi_c\ra$ in the presence of the 
D-brane system is given by:
\be \label{e1}
(Q_B+\bar Q_B)|\Psi_c\ra = |\BB\ra\, ,
\ee
where $Q_B$ and $\bar Q_B$ are the holomorphic and anti-holomorphic 
components of the BRST charge respectively.
Since the closed string field includes the graviton modes, the boundary 
state $|\BB\ra$ contains information about the energy momentum tensor of 
the D-brane system. A simple way to determine the energy momentum tensor 
is to apply $(Q_B+\bar Q_B)$ on both sides of equation \refb{e1}. This 
gives:
\be \label{e2}
(Q_B+\bar Q_B)|\BB\ra = 0\, .
\ee
This equation contains information about the conservation laws of the 
closed string source, in particular of the energy momentum tensor. To see 
how this comes about, let us consider the level (1,1) states\footnote{We 
shall measure level by taking $(c_0 +\bar c_0)
c_1
\bar c_1 |k\ra$ to have level zero.} 
in $|\BB\ra$ 
which are antisymmetric under the exchange of holomorphic and 
anti-holomorphic modes of matter and ghost fields (since the graviton 
state is anti-symmetric under 
such an exchange). The general form of this part of the boundary 
state is given by:
\be \label{e3}
|\BB_0\ra \propto \int d^{26} k [\wt A_{\mu\nu}(k) \alpha^\mu_{-1} 
\bar\alpha^\nu_{-1}
+ \wt B(k) (b_{-1} \bar c_{-1}  + \bar b_{-1} c_{-1})] (c_0 +\bar 
c_0) 
c_1 
\bar c_1 |k\ra\, ,
\ee
where $\wt A_{\mu\nu}=\wt A_{\nu\mu}$, and $\alpha^\mu_n$, $b_n$, $c_n$, 
$\bar\alpha^\mu_n$, $\bar b_n$, $\bar c_n$  are various matter and ghost 
oscillators. The anti-symmetry of the state under left-right exchange 
comes from the anti-symmetry of $c_1\bar c_1$ under such an exchange.
Although we have included integration over all 26 components 
of the momentum, depending on the D-brane configuration, the function 
$\wt A_{\mu\nu}$ and $\wt B$ may have support only over a subspace of the 
26-dimensional momentum space. In writing down \refb{e3} we have also used 
the condition that the boundary state is annihilated by $(b_0-\bar b_0)$ 
and $(c_0+\bar c_0)$.

In the $\alpha'=1$ unit, we have 
\be \label{e4}
(Q_B+\bar Q_B)|\BB_0\ra \propto \sqrt 2\, \int d^{26} k [k^\nu \wt 
A_{\mu\nu}(k) 
+
k_\mu \wt B(k)] 
(\bar c_{-1} \alpha^\mu_{-1} + c_{-1}\bar\alpha^\mu_{-1})
(c_0 +\bar c_0) c_1
\bar c_1 |k\ra\, ,
\ee
If $A_{\mu\nu}(x)$ and $B(x)$ denote the Fourier transforms of $\wt 
A_{\mu\nu}$ and $\wt B$ respectively, then the equation $(Q_B+\bar 
Q_B)|\BB_0\ra=0$ gives us:
\be \label{e5}
\p^\nu (A_{\mu\nu}(x) + \eta_{\mu\nu} B(x) ) = 0\, .
\ee
This shows that we should identify the conserved energy-momentum tensor as
\be \label{e6}
T_{\mu\nu}(x) = K (A_{\mu\nu}(x) + \eta_{\mu\nu} B(x) )\, ,
\ee
where $K$ is an appropriate normalization constant.

We shall now focus on the rolling tachyon solution of ref.\cite{0203211}. 
For definiteness we focus on D-25-brane in flat space-time. 
For this system the boundary state $|\BB\ra$ is given by 
\be \label{e7}
|\BB\ra \propto |\BB\ra_{c=1} \otimes |\BB\ra_{c=25} \otimes 
|\BB\ra_{ghost}\, ,
\ee
Here $|\BB\ra_{c=1}$ is the boundary state associated with the boundary 
CFT involving the 
$X^0$ field, $|\BB\ra_{c=25}$ is the boundary state associated with the 
$c=25$ 
theory describing the flat space-like directions $X^i$ along which we have 
Neumann boundary condition, and $|\BB\ra_{ghost}$ is the boundary state 
associated with the ghost CFT. $|\BB\ra_{c=25}$ and $|\BB\ra_{ghost}$ are 
given by, respectively, 
\be \label{e8}
|\BB\ra_{c=25} \propto \exp\left(-\sum_{n=1}^\infty {1\over 
n}\alpha^i_{-n} \bar 
\alpha^i_{-n} \right) |0\ra\, ,
\ee
and 
\be \label{e9}
|\BB\ra_{ghost} \propto \exp\left(-\sum_{n=1}^\infty (\bar b_{-n} c_{-n} + 
b_{-n} 
\bar c_{-n}) 
\right) (c_0+\bar c_0)c_1\bar c_1 |0\ra\, .
\ee
Finally as shown in 
\cite{0203211}, the relevant part of $|\BB\ra_{c=1}$ is given by:
\be \label{e10}
|\BB\ra_{c=1} \propto [f(X^0(0) + \alpha^0_{-1} \bar \alpha^0_{-1} 
g(X^0(0))]\, |0\ra\, ,
\ee
where
\be \label{e11}
f(x^0)={1\over 1 + e^{x^0} \sin(\tl\pi)} + {1 \over
1 + e^{-x^0} \sin(\tl\pi)} - 1\, ,
\ee
and,
\be \label{e12}
g(x^0) = \cos(2\tl\pi) +1 - f(x^0)\, .
\ee
$\tl$ is a parameter  related to the initial displacement of the tachyon 
field.
Putting these results together, collecting the level 1 states, and 
comparing the resulting expression with \refb{e3}, we get
\be \label{e13}
A_{00}(x) = g(x^0), \qquad A_{ij}(x) = -f(x^0)\delta_{ij}, \qquad B(x) 
= - f(x^0)\, .
\ee
Hence, from \refb{e6}, we have
\be \label{e14}
T_{00} = K(f(x^0) + g(x^0)) = K (\cos(2\tl\pi) +1)\, ,
\ee
and
\be \label{e15}
T_{0i}=0, \qquad T_{ij} = - 2 K f(x^0) \delta_{ij}\, .
\ee
The overall constant $K$ can be fixed by requiring that at $\tl=0$ the 
energy density must agree with the D$p$-brane tension $\TT_p$. This gives
\be \label{edefk}
K = {1\over 2} \, \TT_p\, .
\ee

We see from \refb{e11} that for $0<\tl\le \pi/2$, as $x^0\to\infty$, 
$f(x^0)\to 0$. Thus 
$T_{ij}\to 0$ and the system has zero pressure. On the other hand the 
energy density $T_{00}$ remains costant.

\sectiono{Generalization to Superstrings}  \label{s3}

We now generalize these results to the case of unstable D-branes or 
brane-antibrane system in superstring theory. The boundary 
state for a 
D-brane in superstring theory is defined in the same way as in the case of 
bosonic string. For determining the energy-momentum tensor we can focus 
our attention on the NS-NS sector of the boundary state.
The ghost number and picture number conservation laws on the disk and the 
sphere can 
be used to conclude that the NS-NS sector boundary state $|\BB\ra$ in 
superstring 
theory has the form:
\be \label{e16}
|\BB\ra = \OO |\Omega,k\ra\, .
\ee
Here $\OO$ is a ghost number zero operator constructed from the bosonic 
and fermionic matter 
oscillators $\alpha^\mu_{-n}$, $\bar \alpha^\mu_{-n}$, $\psi^\mu_{-n}$, 
$\bar\psi^\mu_{-n}$, and oscillators of $b$, $c$, $\bar b$, $\bar c$ as 
well as the 
bosonic ghost fields $\beta$, $\gamma$, $\bar \beta$, $\bar \gamma$.
$|\Omega,k\ra$ is the ghost number 3, picture number $-2$ Fock vacuum 
with momentum $k$:
\be \label{e17}
|\Omega,k\ra = (c_0+\bar c_0) c_1 \bar c_1 e^{-\phi(0)} 
e^{-\bar\phi(0)}e^{ik.X(0)}|0\ra\, ,
\ee
where $\phi$, $\bar\phi$ are bosonized ghost fields\cite{FMS}.
With such a $|\BB\ra$, $\la\BB|\psi_c\ra$ will be non-zero for a ghost 
number 3, picture number $-2$ closed string vertex operator $\psi_c$, 
which are the 
correct quantum numbers for getting a non-zero one point function of 
$\psi_c$ on the 
disk. Since $|\Omega\ra$ is odd under both the left and 
the right moving GSO 
projection, and is symmetric under left-right exchange, the general level 
(1/2,1/2) GSO even contribution to the boundary 
state, anti-symmetric under left-right exchange,
is given by:
\be \label{e18}
|\BB_0\ra \propto \int d^{10} k \, \left( \wt A_{\mu\nu}(k) 
\psi^\mu_{-1/2} 
\bar\psi^\nu_{-1/2} 
+ \wt B(k) (\bar\beta_{-1/2} 
\gamma_{-1/2} - \beta_{-1/2} \bar\gamma_{-1/2} )
\right) |\Omega,k\ra\, ,
\ee
where $\wt A_{\mu\nu}=\wt A_{\nu\mu}$.
Requiring $(Q_B+\bar Q_B)|\BB_0\ra=0$ then gives:
\be \label{e19}
k^\mu (\wt A_{\mu\nu} + \eta_{\mu\nu} \wt B) = 0\, .
\ee
As in the case of bosonic string theory, this lets us identify the energy 
momentum tensor $T_{\mu\nu}$ as:
\be \label{e20}
T_{\mu\nu}(x) = K(A_{\mu\nu}(x) + B(x) \eta_{\mu\nu})\, ,
\ee
for some constant $K$. Here $A_{\mu\nu}$ and $B$ are Fourier transforms of 
$\wt A_{\mu\nu}$ and $\wt B$ respectively.

In section \ref{s4} we shall show that for the rolling tachyon solution on 
a D9 brane in type IIA or a 
D9-$\bar{\rm D}$9 brane in type IIB superstring theory, we have:
\be \label{e21}
A_{00}(x) = g(x^0), \qquad A_{ij}(x)=-f(x^0)\delta_{ij}, \qquad B(x) = 
-f(x^0)\, ,
\ee
where
\be \label{e22}
f(x^0) = {1\over 1 + e^{\sqrt 2 x^0} \sin^2(\tl\pi)} + {1 \over
1 + e^{-\sqrt 2 x^0} \sin^2(\tl\pi)} - 1\, ,
\ee
\be \label{e23}
g(x^0) = \cos(2\tl\pi) +1 - f(x^0)\, .
\ee
$\tl$, as usual is the deformation parameter that labels the initial 
position of the tachyon field. Thus we have
\be \label{e24}
T_{00} = K(g(x^0)+ f(x^0)) = K(\cos(2\tl\pi) +1), \qquad T_{ij} = - 2 K 
f(x^0) \delta_{ij}, \qquad T_{0i}=0\, .
\ee
By requiring that at $\tl=0$ $T_{00}$
is equal to the total tension of the brane system, 
the constant $K$ can be determined to be equal to half of the total 
tension of the brane system.

We see from \refb{e22} that for either sign of $\tl$, 
as $x^0\to \infty$, $f(x^0)\to 0$. Thus we see again that in this limit 
$T_{ij}$ and hence the pressure vanishes, whereas the energy density 
remains constant at $K(\cos(2\tl\pi) +1)$.

\sectiono{Boundary State for Rolling Tachyon Solution in Superstrings} 
\label{s4}

{}From general considerations, the NS-NS sector boundary state for the 
rolling tachyon solution in the case of superstring will be given 
by\cite{POLCAI,CLNY,9510161,9701137}:
\be \label{e25}
|\BB\ra = |\BB, +\ra - |\BB, -\ra\, ,
\ee
where
\be \label{e26}
|\BB, \eps\ra \propto |\BB, \eps\ra_{X^0, \psi^0} \otimes |\BB, 
\eps\ra_{\vec X, 
\vec\psi} \otimes |\BB, \eps\ra_{ghost}\, , \qquad \eps=\pm\, .
\ee
Here $(\vec X, \vec\psi)$ stand for the CFT of the spacelike coordinates 
$X^1,\ldots X^9$ and their fermionic partners 
$\psi^1,\ldots\psi^9,\bar\psi^1,\ldots\bar\psi^9$. Of the different 
factors appearing in \refb{e26}, $|\BB, 
\eps\ra_{\vec X,
\vec\psi}$ and $|\BB, \eps\ra_{ghost}$ have the same form as on a static 
D9-brane:
\be \label{e27}
|\BB, \eps\ra_{\vec X,
\vec\psi} \propto \exp\left( -\sum_{n=1}^\infty {1\over n} 
\alpha^j_{-n}\bar\alpha^j_{-n}\right) \exp\left( -i\eps\sum_{n=0}^\infty 
\psi^j_{-n-1/2}\bar\psi^j_{-n-1/2}\right)|0\ra\, ,
\ee
with sum over $j$ running from 1 to 9, and
\ben \label{e28}
|\BB, \eps\ra_{ghost} &\propto& \exp\left(-\sum_{n=1}^\infty (\bar b_{-n} 
c_{-n} + 
b_{-n} \bar c_{-n})
\right) \nonumber \\
&& \times \,  \exp\left(-i\eps \sum_{n=1}^\infty (\bar 
\beta_{-n-1/2} 
\gamma_{-n-1/2} -
\beta_{-n-1/2} \bar \gamma_{-n-1/2})
\right)\, |\Omega\ra\, ,
\een
where
\be \label{e29}
|\Omega\ra = (c_0+\bar c_0) c_1 \bar c_1 e^{-\phi(0)}
e^{-\bar\phi(0)}|0\ra\, .
\ee

We now need to determine $|\BB, \eps\ra_{X^0, \psi^0}$. 
We shall show that the part of $|\BB, \eps\ra_{X^0, \psi^0}$ which 
involves either no oscillators, or involves at most states created from 
pure momenmtum carrying states by the action of 
$\psi^0_{-1/2}\bar\psi^0_{-1/2}$ is proportional to:
\be \label{epred}
(f(X^0(0)) + i\eps\psi^0_{-1/2}\bar\psi^0_{-1/2} g(X^0(0))) |0\ra\, ,
\ee
where $f$ and $g$ are the same functions defined in eqs.\refb{e22}, 
\refb{e23}. 
Thus the net level $(1/2,1/2)$ contribution to the boundary state is given 
by, upto a constant of proportionality,
\be \label{elevhh}
i \left[ \psi^0_{-1/2}\bar\psi^0_{-1/2} g(X^0(0)) - 
\psi^j_{-1/2}\bar\psi^j_{-1/2} f(X^0(0)) - (\bar
\beta_{-1/2}
\gamma_{-1/2} -
\beta_{-1/2} \bar \gamma_{-1/2})  f(X^0(0)) \right] |\Omega\ra\, .
\ee
Comparison of this with 
\refb{e18} immediately 
gives eq.\refb{e21}.

Thus it remains to establish \refb{epred}.
We follow the approach of \cite{0203211} to first study the boundary state 
in the Wick rotated theory $X^0\to iX$, $\psi^0\to i\psi$, $\bar\psi^0\to 
i\bar\psi$ and inverse 
Wick rotate after obtaining the boundary state. We shall not 
attempt to determine this state completely, but only determine the part 
that corresponds to a linear combination of the Fock vaccum states 
$|k\ra$ carrying $X$ momentum $k$, and states of the form 
$\psi_{-1/2}\bar\psi_{-1/2}|k\ra$. These will be sufficient for 
establishing \refb{epred}. 

In the original theory we switch on a tachyon field proportional to 
$\cosh(x^0/\sqrt 2)$.
In the Wick rotated theory this corresponds to 
a tachyon field proportional to $\cos(x/\sqrt 2)$. This gives rise to 
perturbation by a boundary term proportional to the integral of $\psi 
\sin(X/\sqrt 2)\otimes \sigma_1$ where $\sigma_1$ is an appropriate 
Chan-Paton factor\cite{9808141}. We have $X_L=X_R$, $\psi=\bar\psi$ at the 
boundary.
As in \cite{9808141} we fermionize the field $X$ through the relations:
\be \label{ef1}
X \equiv X_L + X_R\, ,
\ee
\be \label{ef2}
e^{i\sqrt 2 X_R} = {1\over \sqrt 2} (\xi + i\eta), \qquad
e^{i\sqrt 2 X_L} = {1\over \sqrt 2} (\bar\xi + i\bar\eta)\, ,
\ee
where $\xi,\eta$ are right-moving Majorana fermions and $\bar\xi$, 
$\bar\eta$ are left moving Majorana fermions.\footnote{In writing 
\refb{ef2} we have ignored the cocycle factors\cite{0003124}. These could 
give rise to additional $\tl$ independent phases in the final answer. 
However, since we shall determine these phases by an independent argument 
anyway, we shall not include the cocycle factors in our analysis.} There 
is a natural SO(3) 
subgroup acting on $\xi$, $\eta$ and $\psi$. We shall for convenience 
identify $\xi$, $\eta$ and $\psi$ axis as 1, 2 and 3 axis respectively.
The boundary perturbation by 
$\psi \sin(X/\sqrt 2)=\psi\sin(\sqrt 2 X_R)$ corresponds to perturbation 
by a 
term 
proportional to integral of $\psi\eta$.\footnote{We could also have chosen 
$\bar\psi\bar\eta$ since at the boundary 
$(\xi,\eta,\psi)=(\bar\xi,\bar\eta,\bar\psi)$.} The effect of this 
perturbation is an SO(3) rotation about the $\xi$ axis. We shall represent 
it by an SU(2) group element:\footnote{Note that the Chan-Paton factor 
$\sigma_1$ does not affect the NS-NS sector closed string vertex operator. 
In 
studying the effect of the deformation on one point function of closed 
string vertex operators, we could keep track of $\sigma_1$ by formally 
including it in the deformation parameter $\tl$. Since the final answer 
will be even in $\tl$, we can formally expand the answer in a power 
series in $\tl$ and then resum the series, and $\sigma_1$ drops out in 
this process.}
\be \label{ef3}
R = \pmatrix{\cos(\pi\tl) & i \sin(\pi\tl)\cr i \sin(\pi\tl) & 
\cos(\pi\tl)}\, .
\ee
This corresponds to an SO(3) rotation angle of $2\pi\tl$.

In order to determine the NS-NS sector boundary state of the perturbed 
theory, we need 
to compute one point function of closed string vertex operators in this 
perturbed theory.\footnote{We could use the result of ref.\cite{9903123} 
and 
reexpress the boundary state in terms of original coordinates, but we 
shall not do that here.} This can be computed by simply rotating the 
closed 
string vertex operator by the SU(2) rotation $R$, and computing the 
resulting one point function in the unperturbed theory.
First we focus on the part involving linear combinations of Fock vacuum. 
Since the boundary perturbation is periodic in $X$ with periodicity 
$2\pi\sqrt 2$, we might expect that the boundary state will be 
a linear combination of states of momentum $n/\sqrt 2$ with integer $n$. 
But the 
perturbation in fact has an additional symmetry $X\to 2\pi/\sqrt 2$, 
$\psi\to -\psi$, $\bar \psi\to -\bar \psi$. Since the fermionic 
oscillators $\psi_{-n}$, $\bar\psi_{-n}$ always appear in pairs in the 
NS-NS sector 
boundary state, and hence is invariant under this transformation, we can 
conclude that the boundary state is in fact built on states carrying 
momentum $n\sqrt 2$ for integer $n$. Thus we can write the 
oscillator free part of the boundary state as:
\be \label{e30}
\sum_n \wt f_n \, |k=n\sqrt 2\ra\, ,
\ee
where the sum over $n$ runs over all integers, 
and 
$\wt f_n$ are coefficients to be determined. 
These are in fact proportional 
to the one point function of the closed string vertex operator $e^{-in 
\sqrt 2 X}$ on the disk. 
In order to carry out this computation we need to study how the SU(2) 
rotation by $R$ affects this vertex operator. To determine this we note 
that as in \cite{9811237,0108238}, the $e^{in\sqrt 2 X_R}$ part of the 
state 
$e^{in\sqrt 2 X}=e^{in\sqrt 2(X_L+X_R)}$ 
transforms in the $(j=|n|,m=n)$ representation of the SU(2) group, 
whereas the part $e^{in\sqrt 2 X_L}$ remains unchanged. Under 
rotation by $R$ this mixes with all the $(j=|n|, m)$ states in the 
representation, which can be represented as states created by combinations 
of $\p X_R$ and $\psi$ oscillators acting on $e^{i\sqrt 2(n 
X_L+mX_R)}(0)|0\ra$. 
However only the term 
proportional to $(j=|n|,m=-n)$ state, represented by the vertex operator 
$e^{in\sqrt 2(X_L-X_R)}$, has a non-zero one point function, since in 
the unperturbed theory $X$ has Neumann boundary condition. The 
coefficient of the state $(j=|n|,m=-n)$ under a rotation by $R$ of 
$(j=|n|,m=n)$ is given by $D^{j=n}_{n,-n}(R)$, where $D^j_{m,m'}(R)$ is 
the spin $j$ representation matrix of $R$ in the $J_z$ eigenbasis. Thus we 
have:
\be \label{e31}
\wt f_n = D^{j=n}_{n,-n}(R) e^{i\ve(n)}\, ,
\ee
where $e^{i\ve(n)}$ is a phase factor, appearing due to the fact that the 
choice of basis in which we compute $D^j_{m,m'}(R)$ may differ from the 
one we need by multiplicative phase factors, and also because we could get 
additional phases from various cocycle factors that we have ignored in 
fermionizing the boson $X$.
We shall use the convention of \cite{9811237} for $D^j_{m,m'}(R)$, which 
gives:
\be \label{e31a}
D^{j=|n|}_{n,-n}(R) = D^{j=|n|}_{-n,n}(R) = (i\sin(\tl\pi))^{2|n|} 
= 
(-1)^n 
\sin^{2|n|}(\tl\pi)\, .
\ee
Thus we get the component of $|\BB,\eps\ra_{X,\psi}$ involving only 
momentum states to be proportional to
\be \label{e32}
\left[1+ \sum_{n=1}^\infty\, (-1)^n
\sin^{2n}(\tl\pi) \{e^{i\ve(n)} e^{i\sqrt 2 n X(0)}
+e^{i\ve(-n)} e^{-i\sqrt 2 n X(0)}\} \right] |0\ra 
\ee
where we have chosen $\ve(0)=0$.
According to \cite{9808141}, 
for $\tl=1/2$ this should reproduce the boundary state 
for an infinite array of D-branes with Dirichlet boundary condition along 
$X$, placed at the zeroes of the tachyon field $T(x)\propto \cos(x/\sqrt 
2)$, {\it i.e.} at $x={2\pi\over \sqrt 2} (n+{1\over 
2})$.\footnote{Actually these are alternatively 
D-branes and anti-D-branes, but the NSNS sector boundary state is 
insensitive to this difference.} It is easy to check that this happens 
provided we choose $\ve(n)=0$.

After inverse Wick rotation $X\to -iX^0$, and setting $\ve(n)=0$, we get 
the 
oscillator free part of the 
boundary state to be proportional to:
\be \label{e33}
f(X^0(0)) |0\ra\, ,
\ee
where 
\ben \label{e34}
f(x^0) &=& 1 + \sum_{n=1}^\infty (-1)^n \sin^{2n}(\tl\pi) (e^{n\sqrt 2 
x^0} 
+ e^{-n\sqrt 2 x^0}) \nonumber \\
&=& {1\over 1 + e^{\sqrt 2 x^0} \sin^2(\tl\pi)} + {1 \over
1 + e^{-\sqrt 2 x^0} \sin^2(\tl\pi)} - 1\, .
\een
This gives eq.\refb{e22}.

We now turn to the computation of $g(x^0)$. Instead of doing a full 
computation, we shall use  a trick, and that is to use eq.\refb{e24} and 
energy conservation to 
conclude that $f(x^0)+g(x^0)$ must be conserved. Thus we must have:
\be \label{e35}
g(x^0) = C - f(x^0)\, ,
\ee
where $C$ is an $x^0$ independent constant to be determined. We compute 
$C$ by examining the power series expansion of $g$ and $f$ in powers of 
$e^{\pm \sqrt 2 x^0}$, and comparing the constant term 
on the two sides of eq.\refb{e35}.

The constant term in $f$ is 1, as seen from eq.\refb{e34}. 
Eq.\refb{e35} then tells us that the constant term in $g$ is given by 
$C-1$. 
{}From eq.\refb{epred} we 
see that the 
coefficient of the $\psi^0_{-1/2} \bar\psi^0_{-1/2} |0\ra$ term 
(relative to the coefficient of $|0\ra$) in the 
expression for $|\BB,\eps\ra_{X^0,\psi^0}$ is given by $i\eps (C-1)$.
Thus after Wick rotation, the coefficient of $\psi_{-1/2} 
\bar\psi_{-1/2} |0\ra$ in $|\BB,\eps\ra_{X,\psi}$ will be $-i\eps (C-1)$. 
Hence
to compute $C$, we need to 
compute the one point function of 
the closed string state $\psi_{-1/2} \bar\psi_{-1/2} |0\ra$ on the disk in 
the perturbed theory and equate it to $-i\eps(C-1)$. In the unperturbed 
theory $|\BB,\eps\ra_{X,\psi}$ is 
proportional to $\exp(-i\eps\psi_{-1/2} \bar\psi_{-1/2})|0\ra$, and hence 
this one point function is given by $-i\eps$. 
Since 
the effect of the perturbation is an SO(3) rotation by angle $2\tl\pi$ on 
the $(\psi,\eta)$ system without changing $\xi$, this 
converts $\psi_{-1/2} \bar\psi_{-1/2} |0\ra$ to 
$(\cos(2\pi\tl) \psi_{-1/2} + \sin(2\pi\tl) \eta_{-1/2}) 
\bar\psi_{-1/2} |0\ra$. Thus we now need to compute the one point 
function of this in the unperturbed theory. Of this the one point function 
of $\eta_{-1/2}
\bar\psi_{-1/2} |0\ra$ vanishes. On the other hand, as already mentioned, 
the one point function 
of $\psi_{-1/2}\bar\psi_{-1/2} |0\ra$ is $-i\eps$. Thus we get
\be \label{ex1}
-i\eps (C-1) = -i\eps \cos(2\pi\tl) \, .
\ee
This gives 
\be \label{ex2}
C = 1 + \cos(2\pi\tl)\, .
\ee
Using eqs.\refb{e35} and \refb{ex2} we recover \refb{e23}.

The tachyon perturbation proportional to $\tl\cosh(x^0/\sqrt 2)$ that we 
have analyzed here represents a system with total energy less than the 
tension of the unstable D-brane. If we want to consider systems with total 
energy larger than that of the unstable D-brane, we need to consider a 
tachyon configuration proportional to $\tl\sinh(x^0/\sqrt 
2)$\cite{0203211}. This can be obtained from the previous background by 
the replacement $\tl\to -i\tl$, $x^0\to x^0+i\pi/\sqrt 2$. Making these 
replacements in \refb{e34} we get
\be \label{enew}
f(x^0)={1\over 1 + e^{\sqrt 2 x^0} \sinh^2(\tl\pi)} + {1 \over
1 + e^{-\sqrt 2 x^0} \sinh^2(\tl\pi)} - 1 
\ee
Thus $f(x^0)$ still vanishes as $x^0\to\infty$ and the system evolves to a 
pressureless gas.

Finally we note that we have focussed our attention only on the NS-NS 
sector of the boundary state since this is the part that carries 
information about the energy momentum tensor. Analysis of the RR sector of 
the boundary state may yield other important informations as in the 
analysis of \cite{0202210}.

\sectiono{Effective Field Theory Analysis
} 
\label{s5}

In this section we shall give a description of the phenomenon observed in 
the earlier sections using an effective field theory.\footnote{I wish to 
thank 
B.~Zwiebach for asking pertinent questions which led to this analysis.}
The effective action 
for the tachyon field $T$ on a D-$p$-brane
that we shall be using is the one proposed in \cite{effective}:
\be \label{ey1}
S\propto - \int d^{p+1} x \, V(T) \sqrt{-\det A}\, ,
\ee
where
\be \label{ey2}
A_{\mu\nu} = \eta_{\mu\nu} + \p_\mu T \p_\nu T\, .
\ee
The tachyon potential $V(T)$ has a maximum at $T=0$ and a minimum at 
$T=T_0$ 
where it vanishes.
The relevant aspects of this model have been discussed in \cite{0009061}.
The energy momentum tensor $T_{\mu\nu}$ obtained from this action is:
\be \label{ey3}
T_{\mu\nu} \propto -V(T) \, \sqrt{-\det A} \, (A^{-1})_{\mu\nu}\, .
\ee
We now restrict to spatially homogeneous time dependent solutions for 
which $\p_i T=0$. Thus we have:
\be \label{ey4}
A_{00}=-1+(\p_0T)^2, \qquad A_{ij}=\delta_{ij}\, \qquad 
\det A=-1+(\p_0T)^2,
\ee
\be \label{ey5}
(A^{-1})_{00} = (-1+(\p_0T)^2)^{-1}, \qquad (A^{-1})_{ij}=\delta_{ij}\, ,
\ee
and hence
\be \label{ey6}
T_{00} \propto V(T) (1-(\p_0T)^2)^{-1/2}, \qquad T_{ij}\propto -V(T) \, 
(1-(\p_0T)^2)^{1/2}\, \delta_{ij}, \qquad T_{i0}=0\, .
\ee
Since $T_{00}$ is conserved, we see from this that for a solution with 
fixed energy, as $T$ approaches the minimum $T_0$ of the potential where 
$V(T_0)=0$, $\p_0 
T$ must approach its critical value 1. This, in turn makes $T_{ij}$ vanish 
in this limit. This is precisely what we have observed in the more 
rigorous boundary state analysis.

Note however that since $\p_0 T$ approaches a finite constant 1, in this 
parametrization the minimum $T_0$ of the potential must be at infinity. 
Otherwise $T$ will approach $T_0$ in a finite time.
 
\medskip

{\bf Acknowledgement}:
I would like to thank B.~Zwiebach for many useful
discussions and comments on the manuscript.
This work was supported in part by a grant 
from the Eberly College 
of Science of the Penn State University.

\end{document}